\documentclass[10pt]{article}

\usepackage{amsmath}    
\usepackage{amssymb}    
\usepackage{graphicx}
\usepackage{cite} 

\usepackage[usenames]{color}
\usepackage{cancel}
\usepackage{hyperref}
\usepackage[normalem]{ulem}

\usepackage[labelfont=bf,labelsep=period,justification=raggedright]{caption}

\makeatletter
\renewcommand{\@biblabel}[1]{\quad#1.}
\makeatother

\date{}

\begin{document}

\begin{flushleft}
{\Large
\textbf{Extortion outperforms generosity in iterated Prisoners' Dilemma}
}
\\
Bin Xu$^{1,2}$,
Yanran Zhou$^{2,3}$,
Jaimie W. Lien$^{4,5}$,
Jie Zheng$^{5}$,
Zhijian Wang$^{2}$
\\
\bf{1} College of Economics, Zhejiang Gongshang University, Hangzhou 310018, China
\\
\bf{2} Experimental Social Science Laboratory, Zhejiang University, Hangzhou 310058, China
\\
\bf{3} College of Economics, Zhejiang University, Hangzhou 310058, China
\\
\bf{4} Department of Decision Sciences and Managerial Economics, The Chinese University of Hong Kong, Hong Kong, China
\\
\bf{5} School of Economics and Management, Tsinghua University, Beijing 100084, China
\\
\end{flushleft}

\section*{Abstract}
Promoting cooperation is an intellectual challenge in the social sciences, for which the iterated Prisoners' Dilemma (IPD) is a fundamental framework. The traditional view that there exists no simple ultimatum strategy whereby one player can unilaterally control the share of the surplus has been challenged by a new class of ¡°zero-determinant¡± (ZD) strategies raised by Press and Dyson. In particular, the extortionate strategies can subdue the opponent and obtain higher scores. However, no empirical evidence has yet been found to support this theoretical finding. In a long-run laboratory experiment of the iterated Prisoners' Dilemma pairing each human subject with a computer co-player, we demonstrate that the  extortionate strategy indeed outperforms the generous strategy against human subjects. Our results show that the extortionate strategy achieves higher scores than the generous strategy, the extortionate strategy promotes the cooperation rate to a similar level as the generous strategy does, and the human subjects' cooperation rates in both the extortionate and generous treatments are increasing over time. While our results imply that the human subjects cared about their earnings as well as fairness or reciprocity, we do observe that subjects learned to become increasingly cooperative over time to increase their own monetary payoffs. Our experiments provide the first laboratory evidence in support of the Press-Dyson theory.

\section*{keywords:} iterated prisoners' dilemma; zero-determinant strategy, evolution of cooperation, evolutionary advantage, game theory

\section*{abbreviations:} IPD, iterated prisoners' dilemma; ZD, zero-determinant; ES, extortionate strategy; GS, generous strategy

\section*{Significance Statement}
(1) The ZD strategies successfully unilaterally enforced a linear relationship between human players' scores and the scores of the computer players playing the ZD strategies, as predicted by Press and Dyson.
(2) The extortionate ZD strategists obtained significantly higher scores than the generous ZD strategists. Half of the extortionate ZD strategists even got scores higher than those resulting from mutual cooperation. (3) The human subjects' cooperation rate against the extortionate ZD strategy is as high as that against the generous ZD strategy. Moreover, the human subjects' cooperation rates increase over time in both treatments. (4) Although human subjects display some degree of conditionally cooperative behaviors, most of them are evolutionary and gradually accede to the extortion in the long run. Consequently, the extortionate ZD strategy outperforms the generous ZD strategy in the long run.
\section{Introduction}

Promoting cooperation under adverse short-term individual incentives is an important social challenge, and the iterated Prisoners' Dilemma (IPD) has been widely studied as the game theoretic framework representing this issue~\cite{rapoport1965prisoner,trivers1971evolution,axelrod1981evolution,axelrod1984evolution,fudenberg1986folk,boyd1987no,kendall2007iterated}. In a one-shot two-person prisoners' dilemma, there are two pure strategies: cooperate (C) and defect (D). Each player receives $R$ if they mutually cooperate; each player receives $P$ if they mutually defect; if one player cooperates and the other defects, the defector receives $T$ and the cooperator receives $S$, where $T > R > P > S$ guarantees that in this game the commonly used solution concept Nash equilibrium, is mutual defection, while $2R>T + S$ implies that mutual cooperation is actually the socially best outcome.

Since the computerized tournaments conducted by Axelrod\cite{axelrod1981evolution,axelrod1984evolution}, kindness and fairness appeared to yield the best chance to promote and sustain cooperation. A substantial number of studies suggest that reciprocity makes mutual cooperation feasible\cite{friedman1971non,Binmore1992278,nowak1993strategy,Wedekind02041996,Duffy2009785,bo2011evolution}, and is a favorable strategy in an evolutionary setting~\cite{molander1985optimal,nowak1992tit,stephens2002discounting,nowak2004emergence}.

Surprisingly, Press and Dyson recently discovered a class of so-called ¡°zero-determinant¡± (ZD) strategies which allow a player to unilaterally enforce a linear relationship between his score and that of his opponent\cite{press2012iterated}. In particular, a subclass of ZD strategies, namely the extortionate strategies, have the potential to guarantee that the extortioner's own surplus exceeds the opponent's surplus by a fixed percentage, by making it optimal for the opponent to cooperate. This means that it is possible to maintain cooperation and pursue self-interest at same time. In addition, an extortioner can earn a score exceeding that from constant mutual cooperation. This new finding by Press and Dyson has also stimulated many researchers to further investigate the performance of ZD strategies in various situations ~\cite{stewart2012extortion,hilbe2013evolution,stewart2013extortion,Plotkin2014extortion,hilbe2014extortion,Hilbe2014multiplayerextortion,Chen201446}.
The key insight of the Press-Dyson theory remains that, if a human being's behavior is developed in an evolutionary manner, he will tend to become more cooperative over time, and if so, the ZD strategist will achieve her maximum possible score by exploiting this cooperative tendency of her opponent~\cite{press2012iterated}.

In spite of these significant developments in the literature, the empirical verification of the theory is rather nontrivial. Up to now, there is only one published study which tests the Press-Dyson theory in a laboratory experiment environment. Hilbe et al~\cite{hilbe2014extortion} provided experimental evidence on performances of different ZD strategies played by computers against human subjects. They specified the ZD strategists to play the extortionate strategy or the generous strategy against human subjects in the context of IPD, in which the extortioner is predicted to earn a higher score based on the Press and Dyson's theory. They find that extortion subdues human players, although generosity turns out to be the more profitable strategy, and furthermore, that the cooperation rate of human co-players against extortionate ZD strategies is only half of that against generous ZD strategies. In other words, generosity appears to be the winning strategy after all. The experimental results by Hilbe et al~\cite{hilbe2014extortion} thus appear to refute the Press-Dyson theory, and has inspired other studies~\cite{bruggeman2015deceit}. This apparent inconsistency between theory and experimental results calls for a closer examination of the discrepancy~\cite{levinezheng2015}.

It is worth noting that in the experiment by Hilbe et al~\cite{hilbe2014extortion}, it was not made known to the human subjects that they were actually playing against a strategy executed by a computer program. Due to this feature of their experimental design, it is arguable that effects of two factors, the ZD strategies themselves and the nature of the other player, cannot be well-distinguished in their results. In the real world, a strategy can be executed by a human individual, but can also be executed by a non-human mechanism. For example, a ZD strategist can be understood as an institution which has its own regularities in generosity or extortion when interacting with human beings. In the view of Darwinism, institutions are competitive and the winning institution (ZD strategist) will be selected based on the performance. To begin to test the theory cleanly, we should explore the performance of the strategy independently of potentially more complicated social influences such as human players' attitudes and intentions towards other human players. In other words, which ZD strategy performs better when playing against human subjects who know that they are facing interaction with a ``strategized" computer instead of a human being.

Another factor that might have led to the inconsistency between the Press-Dyson theory and the experimental result by Hilbe et al~\cite{hilbe2014extortion}, could be insufficient learning opportunities for the human subjects in their experimental design. As is known, ZD strategies are complicated. A ZD strategy is described by the probabilities of cooperation given the four possible outcomes of the previous round: $p=(p_{1} , p_{2} , p_{3} , p_{4})$, where $p_{i}, i\in(1,2,3,4)$ is the probability of cooperation given the previous outcome CC, CD, DC and DD, respectively. It is well understood that a strategy involving uncertainty (and stochastic actions) takes longer time to reveal itself than a pure strategy.  The 60-round sessions implemented in the experiments by Hilbe et al~\cite{hilbe2014extortion},  may not be long enough for people to learn the details of a ZD strategy. Lengthier sessions which can accommodate learning are often more ideal when subjects are facing probabilistic environments~\cite{selten2008stationary,crawford95emt, fudenberglevine1993, fudenberglevine1998book, erev2007learning,xu2013cycle,zhijian2014RPS}.

In order to test the true performance of the ZD strategies, we conducted a laboratory experiment of iterated Prisoners' Dilemma, which provided the human subjects with a rich learning experience over the course of 500 rounds, and made it clearly known to the subjects that they are playing against a strategized computer. Our results show that the extortionate ZD strategy indeed outperforms the generous ZD strategy. To the best of our knowledge, this is the first experimental evidence which supports the predictions of the Press-Dyson theory.

\section{Experiment and Results}
For comparison of performance of different ZD strategies, we designed a laboratory experiment, in which, every human player faced a fixed ZD strategy implemented by a computer program for 500 rounds. Every human player was informed that the opponent will be a computer program.

\begin{figure}[!ht]
\begin{center}
\includegraphics[width=1.5in]{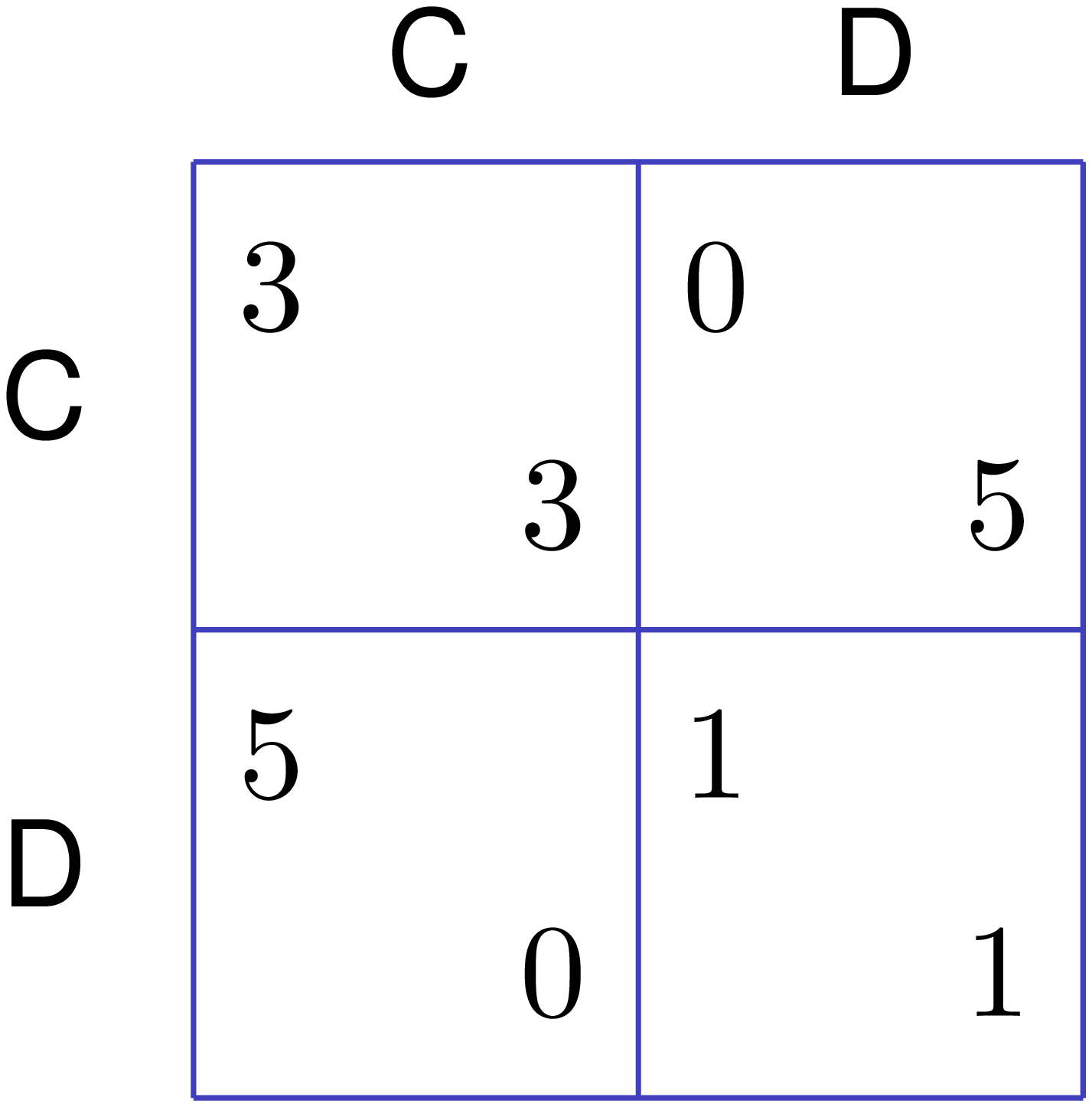}
    \end{center}
  \caption{
    \textbf{Payoff matrix.}
If both players cooperate (C), each player receives 3, if one cooperates and one defects (D), cooperator receives 0 and defecter receives 5, if both defect, each player receives 1.
  }
   \label{payoffmatrix}
\end{figure}

The payoff matrix used in our experiment is shown in Fig~\ref{payoffmatrix} which is the same as in Refs~\cite{press2012iterated,hilbe2014extortion}. Both players receive 3 if they mutually cooperate, both players receive 1 if they mutually defect, the defector receives $5$ and the cooperator receives $0$ if one player cooperates and the other defects.

We employed two treatments, extortionate ZD strategy (ES) and generous ZD strategy (GS). Table~\ref{experimentaldesign} summarizes the experimental design. For ES, the four conditional cooperation probabilities are $(p_{1},p_2,p_3,p_4)=(0.692,0.000,0.538,0.000)$, and for GS the probabilities are $(p_{1},p_2,p_3,p_4)=$ $(1.000,$ $0.182,$ $1.000,$ $0.364)$. In addition, ES defects in the first round and GS cooperates in the first round. These are the same as the strong extortionate ZD strategy and strong generous ZD strategy in Hilbe et al~\cite{hilbe2014extortion}, respectively.

\begin{table}
\caption{\bf{Experimental design}}
\center
\small
\begin{tabular}{|ccccccccc|}
\hline
  Treatment&Number of human co-players &~~ $p_{0}$ ~~&~~$p_{1}$ ~~&~~ $p_{2}$ ~~&~~ $p_{3}$ ~~&~~ $p_{4}$ ~~&  &slope (s) \\
  \hline
  E &32 &0.000& 0.692& 0.000 & 0.538 & 0.000 &  & $1/3$\\
  G &32 &1.000& 1.000 & 0.182 & 1.000 & 0.364 & & $1/3$\\
\hline
\hline
\end{tabular}
\begin{flushleft}
 E, extortion; G, generosity.\\
  ZD strategies are defined by five probabilities: $p_0$, the probability to cooperate in the first round, $p_1$, $p_2$, $p_3$, $p_4$ the four conditional cooperation probabilities given the previous round's outcome CC, CD, DC, and DD, respectively, from the ZD strategist's view. Extortionate strategies do not cooperate in the first round, and they never cooperate after mutual defection. Generous strategies, on the other hand, cooperate in the first round and they always cooperate after mutual cooperation. The parameter s determines the slope of the predicted payoff relationship: a slope of $s=1/3$ implies that for each additional point that the ZD strategist earns, the human co-player's additional payoff is $1/3$.
\end{flushleft}
\label{experimentaldesign}
\end{table}

Altogether 64 graduate and undergraduate students participated in the experiment with each treatment consisting of 32 participants. For further details on the implentation of the experiment, see the \textbf{Materials and Methods} section.
\begin{table}[ht]
\caption{\bf{Summary of the experimental results}
}
\begin{tabular}{@{\vrule height 10.5pt depth 4pt  width0pt}lcccccccccccccccc}
&&\multicolumn5c{Scores per round}&&\multicolumn5c{Cooperation rate}\\
\noalign{\vskip-11pt}
Treatment&number of player\\
\cline{3-13}
\vrule depth 6pt width 0pt &&\multicolumn2c{Human}&&\multicolumn2c{ZD strategist}&&\multicolumn2c{Human}&&\multicolumn2c{ZD strategist}\\
\cline{3-13}
\vrule depth 6pt width 0pt &&\multicolumn1c{$mean$}&$s.d.$&&$mean$&$s.d.$&&$mean$&$s.d.$&&$mean$&$s.d.$\\
\hline
ES &32&1.703&0.200&&2.943&0.511&&0.684&0.198&&0.436&0.137\\
GS &32&2.746&0.260&&2.263&0.788&&0.645&0.349&&0.741&0.245\\
\hline
\end{tabular}\label{experimentalresults}
\end{table}

\subsection{Comparison of the score performance of two ZD strategies}

\begin{figure}[!ht]
\begin{center}
\centerline{  \includegraphics[width=.45\textwidth]{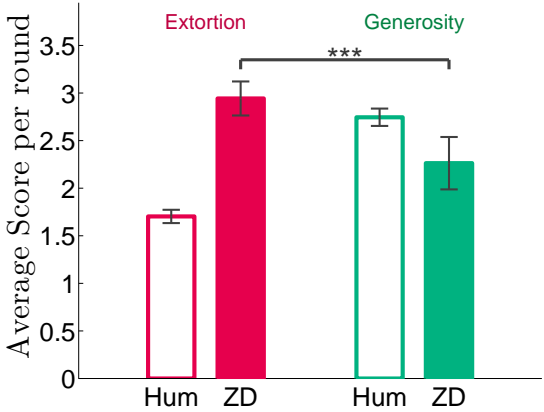}}
    \end{center}
  \caption{
    \textbf{Average scores across the two treatments for human subjects and the ZD strategies.} The scores for the extortionate strategy (red filled bar) are higher than the generous strategy (green filled bar). Three grey stars indicate significance at the level $\alpha=0.001$ (Mann-Whitney test, $n_E=n_G=32$). In terms of average scores, while the extortioners stay ahead of their human co-players (red empty bar), the generous strategists are left behind their human opponents (green empty bar). The error bars indicate the standard errors.}
   \label{fig:Payoffs}
\end{figure}

Table~\ref{experimentalresults} exhibits the overview of the experimental results and Figure~\ref{fig:Payoffs} shows the resulting average scores over all 500 rounds of the game for each of the two treatments. On average, the extortioners earn $2.943\pm0.511$ (mean $\pm$ s.d.) scores per round and generous strategists earn $2.263\pm0.788$ (mean $\pm$ s.d.) scores per round. Extortionate strategists earn significantly higher scores than the generous counterparts (Mann-Whitney test, $n_E=n_G=32$, $z= 4.196$, $p<0.000$). The average scores for extortionate strategists are $30\%$ higher than generous strategists. Furthermore, half of the extortioners even earned scores higher than 3 per round, which is the score associated with constant mutual cooperation. On the contrary, no generous strategist earned scores exceeding 3. This result matched the theoretical prediction of ZD strategies~\cite{press2012iterated} well. For further details, see \textbf{SI}.

\subsection{Comparison of the effectiveness of promoting cooperation between two ZD strategies}
Table~\ref{experimentalresults} also displays the overall cooperation rates. On average, the cooperation rate of human co-players is $0.684\pm0.198$ (mean $\pm$ s.d.) in ES treatment and $0.645\pm0.349$ (mean $\pm$ s.d.) in GE treatment. There is no significant difference between the two treatments (Mann-Whitney test, $n_E=n_G=32$, $z=0.537$, $p=0.591$).

\begin{figure*}[ht]
\begin{center}
\centerline{\includegraphics[width=0.9\textwidth]{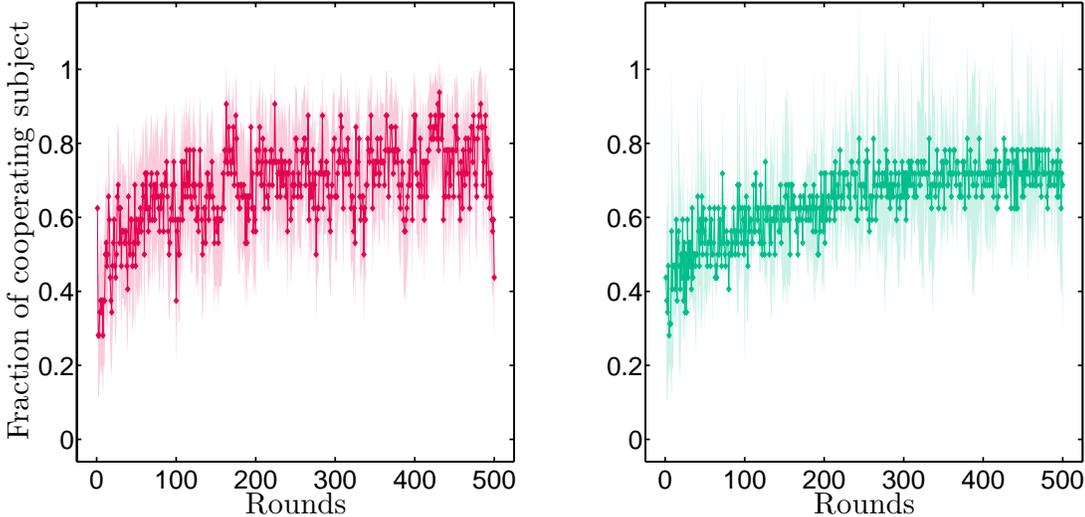}}
\caption{\textbf{Human cooperation rates over the course of the game.} The graph shows the fraction of cooperating human subjects for each round for ES treatment (left) and GS treatment (right). Dots
represent the average human cooperation rate at a round within a treatment, with the shaded areas depicting the $95\%$ confidence interval. The rising trends of cooperation behaviour appear in both treatments (Spearman's rank correlation, for ES, $n=500$, $\rho=0.556$, $p<0.000$; for GS, $n=$500, $\rho=0.757$, $p<0.000$).}\label{fig:Cooperate500rounds}
\end{center}
\end{figure*}

In accordance with the similarity in cooperation rates, the two ZD strategies also generate similar dynamical patterns. Figure~\ref{fig:Cooperate500rounds} shows the dynamical cooperation rate of human subjects for each treatment. The cooperation rate in each treatment starts from a relatively lower point then follows an increasing path. For the ES treatment, compared with the cooperation rate $0.563\pm0.221$ (mean $\pm$ s.d.) in the first 100 rounds, the cooperation rate $0.757\pm0.220$ (mean $\pm$ s.d.) in the last 100 rounds is significantly higher (Wilcoxon signed-rank test, $n=32$, $z =4.207$, $p<0.000$). For the GS treatment, compared with the cooperation rate $0.513\pm0.327$ (mean $\pm$ s.d.) in the first 100 rounds, the cooperation rate $0.716\pm0.386$ (mean $\pm$ s.d.) in the last 100 rounds is also significantly higher (Wilcoxon signed-rank test, $n=32$, $z=3.314$, $p<0.000$).

Our results suggest that the effectiveness of promoting cooperative behaviors is similar between the extortionate strategy and the generous strategy. These results strongly support the theoretical prediction of ZD strategies~\cite{press2012iterated}. A theoretical explanation for the increasing trend in the cooperation rate is provided in \textbf{SI}.

\subsection{The score relationship between ZD strategies and human players}
\begin{figure}[!ht]
\begin{center}
   \includegraphics[width=0.4\textwidth]{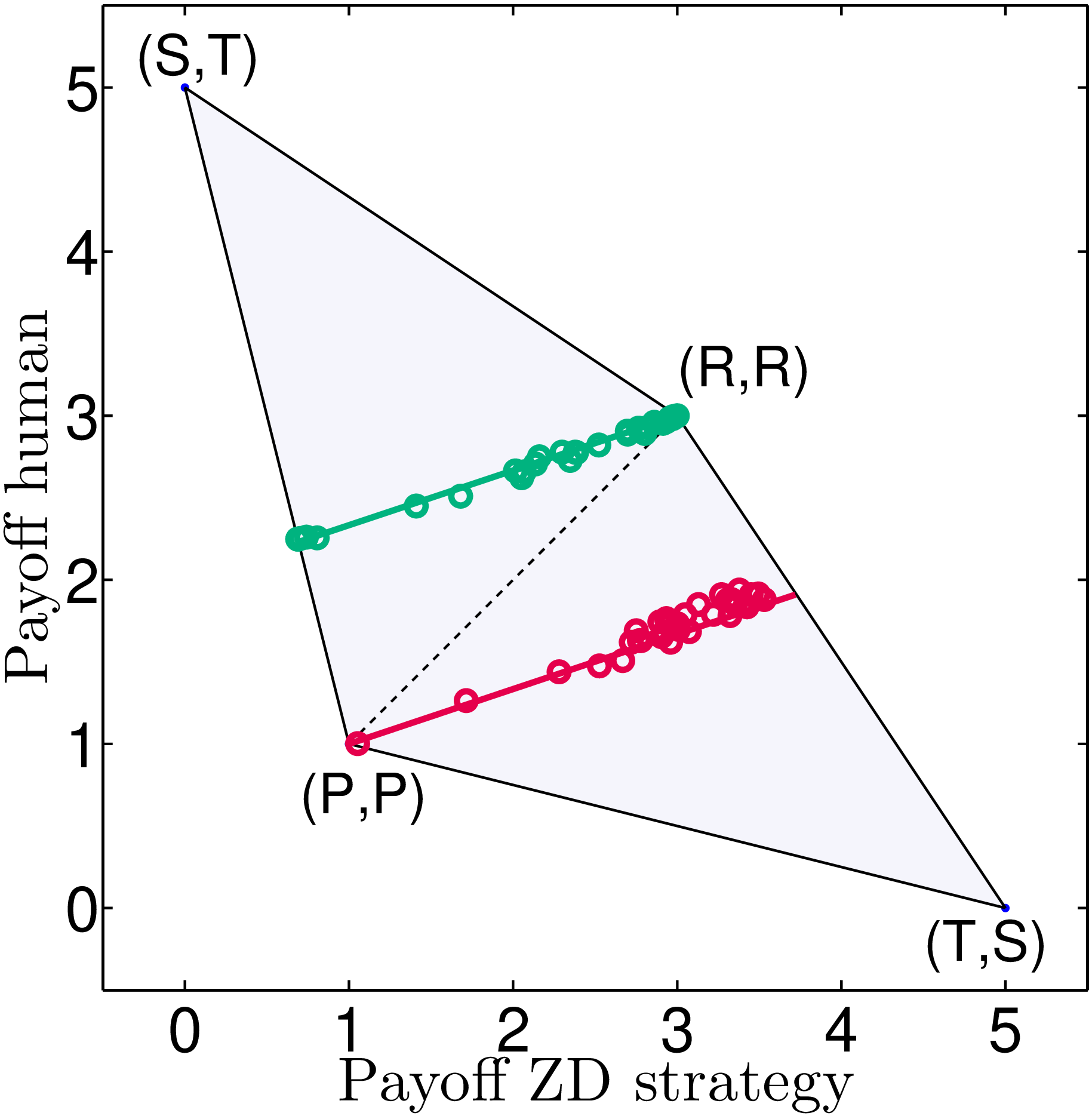}
    \end{center}
  \caption{
    \textbf{Theoretical prediction for the expected scores and experimental performance.}  The grey-shaded area depicts the space of possible scores for the ZD strategy implemented by the computer programme ($x$ axis) and the human co-player ($y$ axis). The red (green) line corresponds to the theoretical prediction for the expected scores of extortion (generosity), respectively, and the open red (green) circles indicate the outcome of the ES (GS) treatment, respectively. Each cycle indicates a pair of a ZD strategy and a human player.}
   \label{fig:Payoffs2}
\end{figure}

Table~\ref{experimentalresults} exhibits the average scores per round for both human subjects and computer programs, and Figure~\ref{fig:Payoffs} illustrates the results. The extortioners earn higher scores than their human co-players (Wilcoxon matched-pairs signed-rank test, $n_{E}=n_{human}=32$, $z=4.937$, $p<0.000$). On the contrary, the generous ZD strategists earn lower scores than their human co-players (Wilcoxon matched-pairs signed-rank test, $n_{G}=n_{human}=32$, $z=4.910$, $p<0.000$). In addition, the relationship of the scores between ZD strategists and human co-players follows the linear relationship prediction, as shown in Figure~\ref{fig:Payoffs2}. These results suggest that ZD strategies can indeed unilaterally enforce a linear relationship between human players' scores and their own scores. These results are consistent with the theoretical prediction of the ZD strategies (for more details, see SI).

\section{Discussion}
Our long-run experiment showed that the extortionate strategy outperformed the generous strategy significantly. The extortionate strategy earned higher scores than the generous strategy while promoting the human co-player's cooperative behavior at the same level of effectiveness. We also found that the ZD strategies was successful in unilaterally enforcing a linear relationship between human players' scores and their own scores. These findings confirm the ZD theory~\cite{press2012iterated} precisely. In particular, half of the extortionate strategists earned scores higher than 3 per round on average, indicating that the extortionate strategist can in fact earn higher payoffs compared to the case of constant mutual cooperation.

Our results are quite different from the previous experimental investigation of Hilbe et al~\cite{hilbe2014extortion}.
In their experiment, the generous strategies earned higher scores than the extortionate strategies, the human co-players' cooperation rate in the extortion treatment is only half of that in the generosity treatment, and the rising trend in the cooperation rate appears only in the generosity treatments. Based on the differences between our results and theirs, we provided evidence to support the ZD theory while their results appeared to raise questions about the theory's empirical validity.

As mentioned above, our experimental design has two key different features from the previous investigation~\cite{hilbe2014extortion}. First, we informed subjects that they are playing against a computer while the same information was not provided to the subjects in the previous investigation. Second, we provided 500 rounds of play while the previous study provided 60 rounds of play. Human subjects may need lengthy periods of time in order to understand the decision problem or strategic issue they face (see Chapter 1 in~\cite{camerer2003behavioral}).

As the previous investigation did not make the nature of the opponent (being a computer or a human being) apparent to human subjects~\cite{hilbe2014extortion}, the authors fully ascribed the inconsistency between the theoretical predictions and the experimental results to a preference for conditional cooperation~\cite{fischbacher2001people,fischbacher2010social,chaudhuri2006conditional} or fairness~\cite{fehr1999theory,brosnan2003monkeys,oechssler2013finitely}. Existing literature has found that human individuals in general show and acquire a stronger sense of fairness when they are playing against other human individuals compared to playing against computer programs~\cite{blount1995social,gintis2003solving}. Hilbe et al show that people cooperate more when the opponent cooperated in the previous round~\cite{hilbe2014extortion}. Human subjects in their study would defect when they played against the extortionate strategies, because the extortioner often defected. We also found the propensity of conditional cooperation in our experiment where the human subjects knew for sure that they faced a computer program. Human subjects' cooperation rates were higher if the ZD strategists cooperated in the previous round than otherwise. On average, human cooperation rates were $78.45\%$ if the extortionate strategist cooperated in the previous round, and $60.83\%$ otherwise, and $66.49\%$ if the generous strategist cooperated in the previous round, and $41.20\%$ otherwise. However, in our experiment, we noticed that even if the opponent defected in the previous round, the cooperation rates were still high, especially in the ES treatment. This cannot be explained by conditional cooperation. The requirement for the human co-player to earn high scores is precisely to cooperate unconditionally.

We also observed that the increasing cooperation trend lasted for half of the 500 rounds in both treatments. It involves a learning process for the human subjects to realize that cooperation is beneficial. In the experiment, a human subject might have found that, for each of the four possible previous outcomes CC, CD, DC and DD, she/he had two choices C and D, then she/he could try to figure out in total eight conditional scores. Theoretically, the results are shown in Table~\ref{experimentaldesign2}, given that a ZD strategist has a unique conditional strategy $p_1$, $p_2$, $p_3$, $p_4$. For human subjects, if any of the possible outcomes in the previous outcome are considered, strategy D forever dominates strategy C in both treatments. This might be the reason that the cooperation rates during early stages are relatively lower. However, as time goes on, an intelligent and careful player would find that the lowest scores in the first two conditions are higher than the highest scores in the second two conditions in both treatments (a theoretical analysis will be provided in \textbf{SI}), and the right choice is C. Thus, the cooperation rates followed a gradually increasing process. The result also indicates that one may not be able to find evidence to support the ZD theory~\cite{press2012iterated} if the number of rounds of play is not sufficiently large.

\begin{table}[h]
\caption{Human's expected scores of conditional strategies.
}
\center
\begin{tabular}{@{\vrule height 10.5pt depth 4pt  width0pt}cccccc}
\multicolumn{1}{l}{}&\multicolumn1c{}&\multicolumn1c{CC}&
\multicolumn1c{CD}&\multicolumn1c{DC}& \multicolumn1c{DD}\cr
\cline{3-6}
\multicolumn{1}{l}{Treatment}&\multicolumn1c{Strategy}&\multicolumn1c{$p_1$}&
\multicolumn1c{$p_3$}&\multicolumn1c{$p_2$}& \multicolumn1c{$p_4$}\\
\hline
\multicolumn{1}{l}{}&\multicolumn{1}{c}{C}&\multicolumn{1}{c}{2.076} &{1.614}&0.000&0.000\cr
\multicolumn{1}{l}{ES}&\multicolumn{1}{c}{D}&\multicolumn{1}{c}{3.768} &3.152&1.000&1.000\\
\hline
\multicolumn{1}{l}{}&\multicolumn{1}{c}{C}&\multicolumn{1}{c}{3.000} &3.000&0.546&1.092\\
\multicolumn{1}{l}{GS}&\multicolumn{1}{c}{D}&\multicolumn{1}{c}{5.000} &5.000&1.728&2.456\\
\hline
\end{tabular}\label{experimentaldesign2}
\begin{flushleft}
CC, CD, DC and DD are the previous outcomes from the human subject's view. For the details of the calculation, see SI.
\end{flushleft}
\end{table}

In economics, a rational agent is assumed to maximize her/his own benefit, and the ZD strategies have the feature of ensuring that the opponent's best response is to fully cooperate. While our results suggest that the human subjects cared about their earnings as well as fairness or reciprocity, we do observe that subjects learned to become increasingly cooperative over time to increase their own monetary payoffs. Evolution has an essential impact on both human subjects and the ZD strategist. As predicted by Press-Dyson theory, in our laboratory experiments we have observed human behavior evolving towards full cooperation, while the extortionate strategy outperforms generosity and wins by social selection.

\section{Materials and Methods}\label{Methods}
\subsection{Data source and Experimental setting}
The data we used here come from our laboratory experiments which were conducted at The Experimental Social Science Laboratory of Zhejiang University, on 19 and 21 November 2014. A total of 64 undergraduate and graduate students from various disciplines were recruited to participate in the experiment with each of them only participating once. In total, we collected 64000 observations of individual decision making, consisting of choices of human subjects and ZD strategies implemented by computer programs.

There were two treatments in total, one was extortionate ZD strategy (ES) and the other was generous ZD strategy (GS). For each treatment, there were 32 participants with half of them being male and the other half being female. Each experimental session consisted of 4 pairs of human players and computers, and in total there were 16 sessions in the set of experiments. On average, each session lasted for 1 hour. During the experiment, the player earned scores according to the payoff matrix (see Figure~\ref{payoffmatrix}) and their choices. After the experiment, the sum of scores were converted to cash according to an exchange rate and paid to the subjects. The average earning is about 50 Yuan RMB including the 5 Yuan show-up fee.

Before the formal experiment, the subjects practiced with a matching pennies game against a computer to get acquainted with the laboratory setting. They were then assigned a set of materials including an instruction manual (see SI), an informed consent form and a recording chart for their use, and they played the game in an small isolated room with a computer. Oral instructions were also given. They made decisions by clicking the option C or D on the screen. The software for the experiment was designed by the authors. No communication was allowed, and the subjects were asked to put their mobile phones in mute and sealed in an envelope until the end of the session. Considering the complexity of the ZD strategies, we provided human subjects with paper and pen to record their decision choices and scores round by round. The human subjects were told that they would play a game with a fixed computer program for more than 500 rounds, and after the 500th round, the game will end after each round with a probability of 0.1. This setting allows us to get 500 rounds data for each subject while avoiding end-of-game effects~\cite{selten1986end}.

\subsection{Statistical methods}
Throughout the paper, we use two-tailed non-parametric tests for our statistical analysis. With each iterated game between a human co-player and the computer as our statistical unit, we have 32
independent observations for each of the two treatments and each of two players (a computer and a human player). Specifically, we used Mann-Whitney test for the comparison between treatments, Wilcoxon signed-rank test for the comparison within a treatment. In addition, we use Spearman¡¯s rank correlation test for trend test.

\section*{Acknowledgments}
We thank Anping Sun for excellent assistance. This work was partially supported by the Fundamental Research Funds for the Central Universities (SSEYI2014Z), and partially supported by the Philosophy and Social Sciences Planning Project of Zhejiang Province (13NDJC095YB).


\end{document}